\documentclass{IEEEtran}

\usepackage{graphics,amssymb,array,cite,bm}
\usepackage[active]{srcltx}
\usepackage[english]{babel}
\usepackage[dvips]{graphicx}
\usepackage{hyperref}
\usepackage{amsfonts}
\usepackage{bbm}
\usepackage{mathrsfs}
\usepackage{xspace}
\usepackage{bm}

\usepackage{color}
\usepackage[cmex10]{amsmath}

\usepackage{algpseudocode,algorithm,algorithmicx}
\usepackage{enumitem}

\newcommand{\eeq}{\end{equation}}

\newcommand{\beq}{\begin{equation}}

\begin{document}

\title{Model-Aided Wireless Artificial Intelligence:\\ Embedding Expert Knowledge in Deep Neural Networks Towards Wireless Systems Optimization}
\author{{Alessio Zappone, {\em Senior Member, IEEE}, Marco Di Renzo, {\em Senior Member, IEEE}, M\'erouane Debbah, {\em Fellow, IEEE}, Thanh Tu Lam, and Xuewen Qian}

\thanks{The authors are with the Laboratory of Signals and Systems (CNRS - CentraleSupelec - Univ. Paris-Sud), Universit\'e Paris-Saclay, 3 rue Joliot-Curie, 91192 Gif-sur-Yvette, France, (alessio.zappone@l2s.centralesupelec.fr, marco.direnzo@l2s.centralesupelec.fr).}


}
\maketitle

\begin{abstract}
Deep learning based on artificial neural networks
is a powerful machine learning method that, in the last few
years, has been successfully used to realize tasks, e.g., image
classification, speech recognition, translation of languages, etc.,
that are usually simple to execute by human beings but extremely
difficult to perform by machines. This is one of the reasons why
deep learning is considered to be one of the main enablers to
realize the notion of artificial intelligence. In order to identify the
best architecture of an artificial neural network that allows one
to fit input-output data pairs, the current methodology in deep
learning methods consists of employing a data-driven approach.
Once the artificial neural network is trained, it is capable of
responding to never-observed inputs by providing one with the
optimum output based on past acquired knowledge. In this
context, a recent trend in the deep learning community is to
complement pure data-driven approaches with prior information
based on expert knowledge. In this work, we describe two methods that implement this strategy, which aim at optimizing wireless
communication networks. In addition, we illustrate numerical
results in order to assess the performance of the proposed
approaches compared with pure data-driven implementations.
\end{abstract}

\section{Introduction and Motivation}
Traditional deep learning approaches consist of acquiring a
large amount of empirical data about the system behavior and
employing it for performance optimization. We believe, on the
other hand, that the application of deep learning to communication networks design and optimization offers more possibilities than such a pure data-driven approach. As opposed
to other fields of science, such as image classification and
speech recognition, mathematical models for communication
networks optimization are very often available, even though
they may be simplified and inaccurate. We believe that this a
priori expert knowledge, which has been acquired over decades
of intense research, should not be dismissed and ignored. Thus,
this work puts forth a new approach that capitalizes on the
availability of (possibly simplified or inaccurate) theoretical
models, in order to reduce the amount of empirical data to
use and the complexity of training artificial neural networks
(ANNs). We show that the enabling tool to realize this vision
is transfer learning. Unlike other application areas, we provide numerical evidence that synergistically combining prior
expert knowledge based on analytical models and data-driven methods constitutes a suitable approach towards the design and
optimization of communication systems and networks with the
aid of deep learning based on ANNs.

\section{Artificial Intelligence by Deep Learning}\label{Sec:I}
Artificial intelligence (AI) through machine learning enables
machines to perform tasks by learning from data, instead
of running a static computer program. In this context, deep
learning is a specific machine learning method that implements
the learning process by employing ANNs \cite{Bengio2016}. In principle,
ANNs are able to operate in a fully data-driven fashion, thus
dispensing system designers with the need of mathematical
modeling and expert supervision. When large
datasets are available, moreover, deep learning is known to
outperform other machine learning techniques. These features
have made deep learning the most widely used machine learning technique in fields such as image classification, speech
recognition, translation between languages, autonomous driving, etc., for which a mathematical description of the task to
be executed is particularly challenging to obtain. In these areas
of research, however, an emerging opinion is that pure data-driven approaches may become unfeasible in the context of
large-scale applications, due to the huge amount of required
data, and to the related processing complexity. In \cite{Inoue2017}, for
example, image processing for object position detection in
robotic applications is considered. This work observes that augmenting a small training dataset of real images
with a large dataset of synthetic images significantly improves
the estimation accuracy with respect to processing only the
small dataset of real images. Similar results have been obtained
in \cite{Kim_AcousticDNN} for application to speech recognition. These promising
results have motivated us to investigate the interplay of data-driven (i.e., model-free) and model-based methods in wireless
networks.

\section{Deep Learning in Communications: Why Now?}
General machine learning techniques are not new to wireless
communications \cite{SimeoneNowPub}, even though the use of deep learning
has never been considered in the past. In our opinion, this is
mainly due to the fact that, unlike other fields of science where
theoretical modeling is particularly difficult to be performed
and a data-driven approach is often the only solution available, wireless communications have always relied on strong
theoretical models for system design and optimization. This status quo, however, is rapidly changing, and very recently
the use of deep learning has started being envisioned for
wireless communication applications. Indeed, the increasing
complexity of wireless networks makes it harder and harder
to come up with theoretical models that are at the same time
accurate and tractable. The rising complexity of beyond 5G
and future 6G networks is, however, exceeding the modeling
and optimization possibilities of standard mathematical tools,
as recently discussed in \cite{TCOM_AItutorial} and \cite{EURASIP_RIS}.

In addition, the use of deep learning for application to
wireless communications provides communication theorists
and engineers with another major opportunity. As discussed
in Section \ref{Sec:I}, an emerging research trend in the deep learning
community is the development of techniques that exploit
prior information that is available on the problem to solve.
This represents a great opportunity in the context of wireless
communications, because theoretical models, despite their
possible inaccuracy or cumbersomeness, are often available
and provide one with much deeper prior information compared
with other fields of science. This clear advantage of wireless
communications should not be wasted. Accordingly, the aim
of this work is to corroborate the intuition that available
theoretical models and frameworks can indeed provide enough
expert knowledge to facilitate the use of deep learning for
application to wireless networks design. To this end, two main
methods for embedding expert knowledge into deep learning
techniques are discussed, and three specific case studies are
analyzed.

\section{Learning to Optimize by Deep Neural Networks}
A fundamental component of wireless networks management and operation is the allocation of the available resources
to optimize desired performance functions, in order to ensure
guaranteed performance to each user. Depending on the complexity of the system, four major case studies usually arise:
\begin{itemize}
\item \textbf{Case 1}: An accurate and tractable theoretical model is
available (e.g., point-to-point channel capacity, point-to-point bit error probability).
\item \textbf{Case 2}: An accurate but intractable theoretical model is
available (e.g., achievable sum-rate in interference-limited
systems). 
\item \textbf{Case 3}: A tractable but inaccurate theoretical model
is available (e.g., spectral / energy efficiency of ultra-dense networks, energy consumption models, hardware
impairments).
\item \textbf{Case 4}: Only inaccurate and intractable theoretical models are available (e.g., molecular communication networks, optical systems, end-to-end networks optimization).
\end{itemize}

This work will focus on Cases 2 and 3, which offer the best
opportunities of cross-fertilization between model-aided and
data-driven approaches, as discussed in the next two sections.
Moreover, the availability of theoretical models, although
possibly inaccurate or intractable, represents the most common
scenario in wireless communications. On the other hand,
Cases 1 and 4 can be handled by traditional (model-based)
system design approaches and pure data-driven techniques,
respectively.

\subsection{Learning to Optimize a Model}\label{Sec:OptModel}
In Case 2, a mathematical formula for the performance
metric to optimize is available, but it is too complex to be
maximized by using traditional optimization theory methods.
Then, the issue is not solving the optimization problem, but
rather the complexity and the time required to do so, which
might not be compatible with real-time designs of mobile
scenarios, wherein the network status changes rapidly (e.g., a
user joins/leaves the network, the channel realizations/statistics
change, new traffic requests occur, etc.) and thus the optimal
resource allocation policy to use needs to be updated each time
that the network scenario changes (so, very often).

In this context, the joint use of deep learning and traditional
optimization theory can significantly speed up the computation
of the optimal resources. Our proposed key idea is based on
two observations:
\begin{itemize}
\item Resource allocation can be regarded as the problem of
determining the map between the system parameters (e.g.,
the propagation channels, the number of active users,
the users’ positions, etc.) and the corresponding optimal
resource allocation to use.
\item ANNs are known to be universal function approximators,
i.e., they can be trained to learn, virtually, any input-output map \cite{Hornik1989}.
\end{itemize}

Thus, it is possible to consider an ANN whose inputs are
the system parameters and whose outputs are the resources to
allocate, so that its input-output relationship approximates the
map between the system parameters and the optimal resource
allocation policy. Once the ANN is configured, it is possible to
obtain the optimal resource allocation policy without having to
solve any optimization problems in real-time, i.e., every time
that the system configuration changes.

In order for the ANN to learn the desired input-output
map, the ANN needs to be trained in a supervised fashion
by means of a training set, i.e., a dataset containing examples
of system configurations with the corresponding desired (i.e.,
optimal) resource allocation policy. The training process is
accomplished by efficient, off-the-shelf, stochastic gradient
descent algorithms that adjust the ANN parameters in order
to reduce the error between the actual output and the desired
training output. Generating the training set is, on the other
hand, more computationally intensive, because it requires to
solve the resource optimization problem by using (traditional)
optimization techniques. However, the following observations
support the use of ANNs for resource allocation problems:
\begin{itemize}
\item The training set can be generated off-line. Thus, a much
higher complexity can be afforded and real-time constraints does not constitute an issue for this phase.
\item The training set can be updated at a much longer timescale, as opposed to the rate of change of the network
parameters.
\end{itemize}

\begin{figure}[!t]
  \includegraphics[width=\columnwidth]{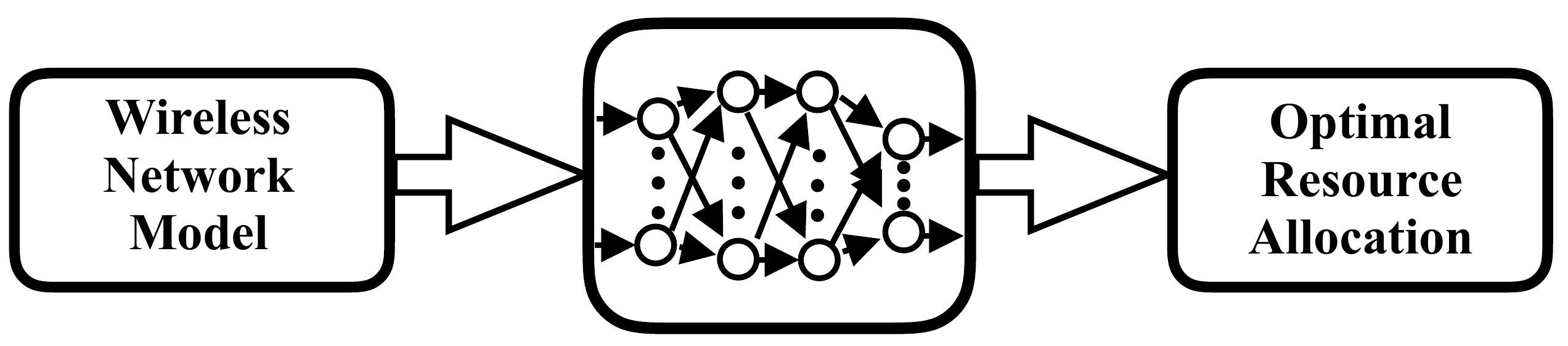}
  \caption{A training set is built from a mathematical model and used to train an ANN that provides one with optimal resource allocation policies.}
  \label{Fig:CaseC2}
\end{figure}

The approach described in this section is schematically
depicted in Fig. \ref{Fig:CaseC2}. In a nutshell, an available accurate model is used for efficiently training the ANN, which, once appropriately trained, can then be used for the efficient implementation
of real-time resource allocation strategies.

\subsection{Learning to Refine a Model}\label{Sec:OptData}
In Case 3, a tractable model is available but it is not
sufficiently accurate. Nevertheless, even inaccurate models can
provide system designers with useful information that should
not be dismissed. Employing a fully data-driven approach to
train an ANN requires, in general, the need of acquiring a huge
amount of live data. This task, however, might not be practical
due to the time, the complexity, or the economical reasons that
the process of acquiring data entails. Instead, the availability
of an approximate model can be exploited to perform a first
rough training of the ANN, which can be subsequently refined
only through a small set of real data. This approach is likely
to reduce the amount of live data to use for training ANNs.

More precisely, we propose the following approach:
\begin{itemize}
\item First, a training set based on a (possibly approximate)
model is obtained, by using conventional optimization
theory, as described in Section \ref{Sec:OptModel}.
\item Then, an ANN is trained by processing the generated
training set, as discussed in Section \ref{Sec:OptModel}. This provides
one with an initial configuration for the ANN.
\item Finally, the pre-trained ANN architecture is refined
through a new training phase based on real/measured
data (i.e., input-output pairs corresponding to the optimal
network configuration).
\end{itemize}

\begin{figure}[!t]
  \includegraphics[width=\columnwidth]{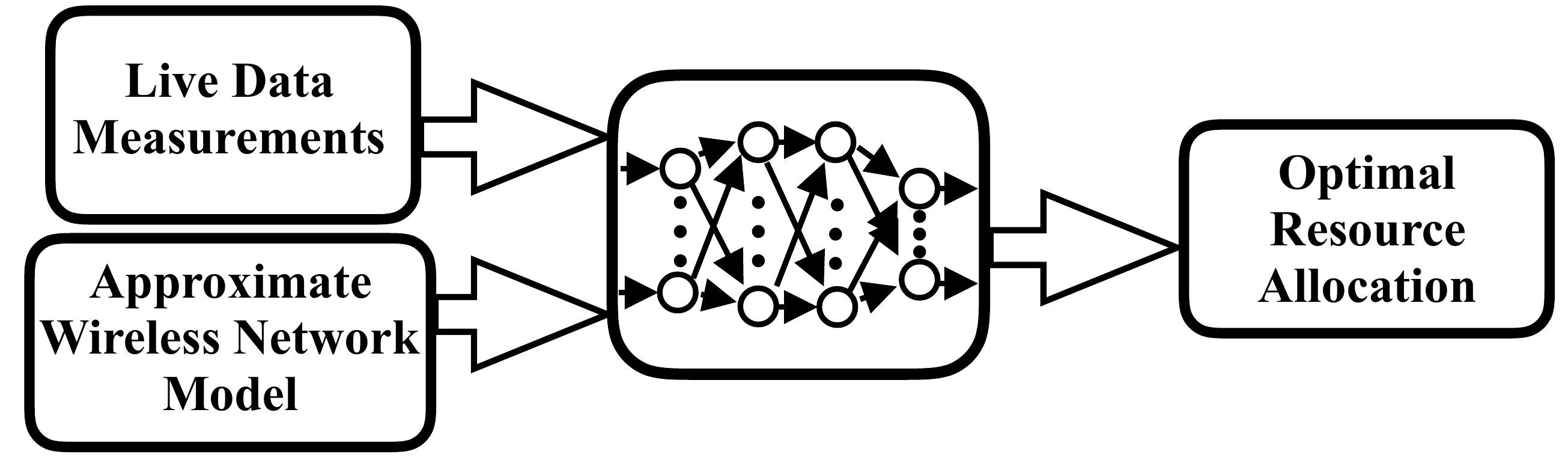}
  \caption{A training set is built from a mathematical model and used to train an ANN that provides one with optimal resource allocation policies.}
  \label{Fig:CaseC3}
\end{figure}

Accordingly, the first training phase provides one with an
initialization point for the second training phase, which has the
potential of significantly reducing the amount of real data to
be acquired compared with the case when no initial guess for the parameters of an
ANN is available. This approach is schematically depicted in
Fig. \ref{Fig:CaseC3}.

The relevance and potential of the approach depicted in Fig. \ref{Fig:CaseC3} cannot be underestimated. It is known, in fact, that selecting
an efficient initialization point when training an ANN is a
critical issue that significantly affects the training process. To
date, only heuristic methods or random initialization strategies
have been proposed \cite{Saxe2013,Sussillo2015}. The approached proposed in Fig. \ref{Fig:CaseC3} can, on the other hand, be viewed as an initialization method
that provides one with a stronger theoretical justification, since
it capitalizes on the prior expert knowledge at our disposal.
Clearly, the performance of the proposed approach strongly
depends on the accuracy of the model that is used to perform
the first training. If the difference between the model and the
real system is not too large, and the initial training phase is performed on enough data, then the second training phase
will start from an initial configuration of the ANN that is
already close to the desired (final) configuration. It is expected,
therefore, that only a small dataset and a few gradient iterations
might suffice to refine the configuration of the ANN from the
initial setup. If, on the other hand, the model is a too inaccurate
approximation of the true system or a small training set is used,
the initial training might yield a misleading configuration of
the final ANN. In some cases, this may require even more
data during the second training phase in order to obtain a
good solution. The optimization of this approach, which is
an instance of transfer learning, is, therefore, a non-trivial
task, especially in the context of optimizing wireless networks
where it has never been applied before. In order to evaluate the
closeness between model-based and live data, several distance
measures between sets can be employed, such as the relative
mean squared error between the datasets. This is elaborated in
the next section.

\section{Application Examples}\label{Sec:Applications}
In this section, we consider three case studies in order to
substantiate the proposed model-aided approaches for training
ANNs. Our objective is two-fold. First, we show that, by using
the approach introduced in Section \ref{Sec:OptModel}, an ANN can be
used to implement real-time resource allocation policies that
are too complex to be handled in real-time by conventional
optimization-theoretic approaches. Then, based on the methodology introduced in Section \ref{Sec:OptData}, we show that an ANN can
be first roughly trained by using (large) datasets based on
analytical, but inaccurate, models, and can be subsequently
finely-tuned by employing datasets of live data with limited
size compared with what would be required if the initial training based on analytical models is not performed. Preliminary,
non-optimized, results were reported in the tutorial paper \cite{TCOM_AItutorial}.
In the present paper, compared with \cite{TCOM_AItutorial}, we comprehensively
study the training and validation performance as a function
of the training epochs \cite{Bengio2016}, and optimize the architecture of
the ANNs accordingly. Thanks to this comprehensive study
and fine optimization tuning, we identify better architectures
for the resulting ANNs, which yield lower test errors. This
is a major innovation compared with \cite{TCOM_AItutorial}, since the proposed
approach ensures that the resulting ANNs are not operating
either in the overfitting or in the underfitting regimes.

In all considered case-studies, ANN training has been performed by employing the ADAM training algorithm implementation by the Keras environment, with Tensorflow backend.

\subsection{Real-Time Energy Efficiency Maximization in Multi-User Networks}
Consider the uplink of a multi-user network with interfering
mobile terminals. The objective is to allocate the transmit
powers of the users in order to maximize the network bit-per-Joule energy efficiency, which is defined as the ratio between
the system sum achievable rate and the total network power
consumption. In this scenario, a model to formulate the energy
efficiency optimization problem is available, but the presence
of multi-user interference makes it too complex to be globally
solved at an affordable computational complexity \cite{ZapNow15}. This
is especially problematic when the optimization is performed
based on instantaneous channel realizations, and thus needs to
be performed anew every time the channel coefficients change.
This causes a considerable complexity overhead that prevents
one from using real-time implementations. The problem can be
overcome by using the approach introduced in Section \ref{Sec:OptModel}. We illustrate it by using a simple example.

A circular area of radius $500\,\textrm{m}$ and with $10$ mutually interfering mobile users is considered. In order to build a training
set, 10,000 independent network scenarios are generated by
randomly dropping the users in the service area, and by modeling the propagation losses according to \cite{PathLossModel} and the fading
channels as standard complex Gaussian random variables. For
each scenario, the optimal energy-efficient power allocation
strategy is computed off-line by using fractional programming \cite{ZapTSP17}. Accordingly, 10,000 training samples are obtained.

A feedforward ANN with rectified linear unit activation (ReLU)
functions and 10 layers is considered. Layers 1 and 2 have
18 neurons, and the number of neurons of the other layers
decreases by two every two layers. Thus, the output layer has
10 neurons and yield the users’ transmit powers.

\begin{figure}[!t]
  \begin{center}
  \includegraphics[width=\columnwidth]{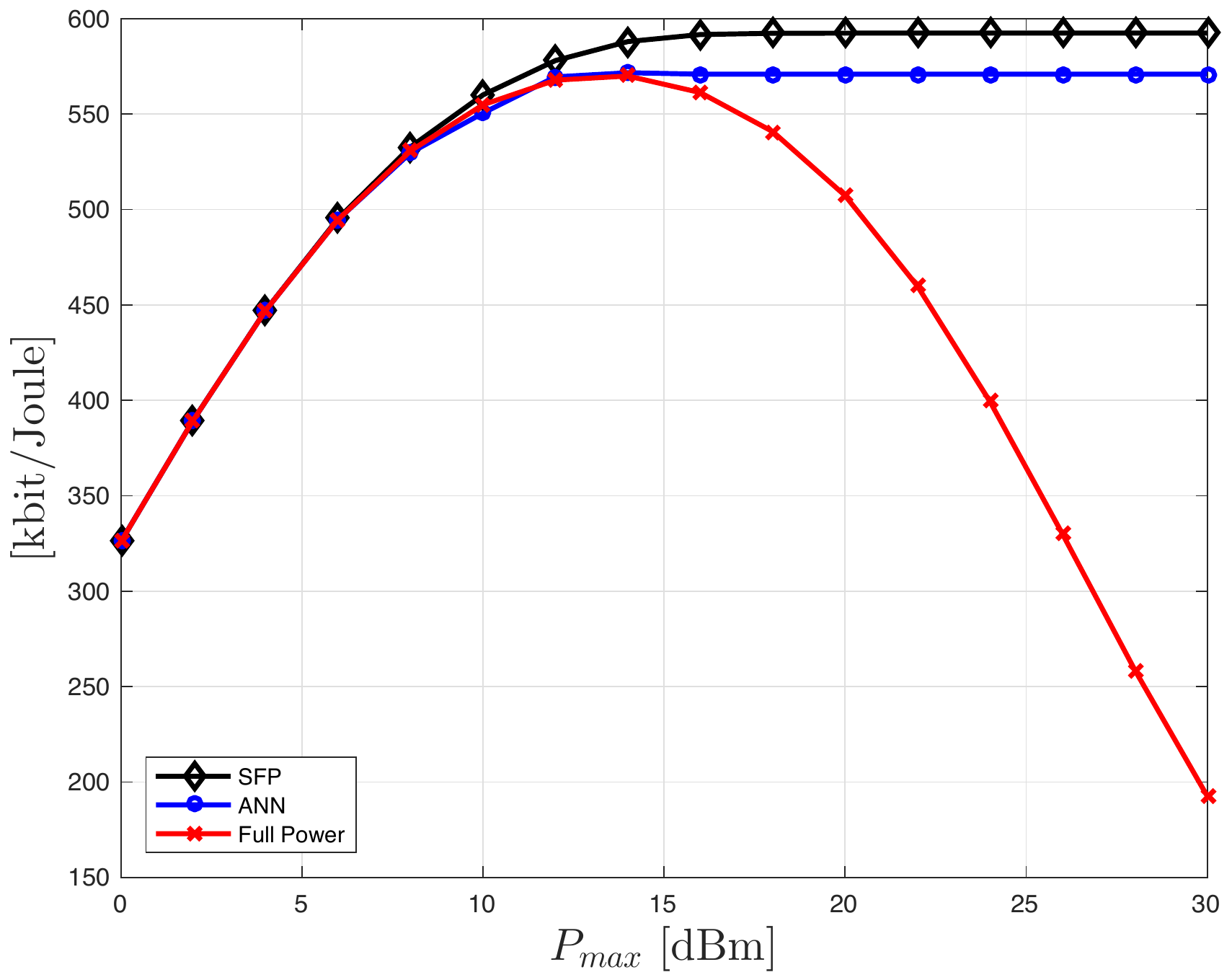}
  \caption{Energy efficiency versus $\rm{P_{max}}$: (black line) Method from \cite{ZapTSP17}; (blue line) Deep learning by using ANN; (red line) Full power allocation.}
  \label{Fig:GEE}
 \end{center}
\end{figure}

The performance of the trained ANN is evaluated over a
test set of 10,000 new channel scenarios, which are generated
by using the same procedure used for the training set, but
with independent users’ drops and channel realizations. Fig. \ref{Fig:GEE} compares the average (over the test set) energy efficiency
obtained using the trained ANN, and the global optimum
obtained via fractional programming, versus the maximum
feasible transmit power  $\rm{P_{max}}$. The energy efficiency obtained
when all the users transmit with power $\rm{P_{max}}$ is reported as
well. It is observed that, despite the much lower complexity,
the ANN-based scheme is optimal for low $\rm{P_{max}}$, and near-optimal for larger $\rm{P_{max}}$, where it achieves around $95\%$ of the optimal value.

\subsection{Energy Efficiency Maximization in non-Poisson Cellular Networks}\label{Sec:CaseII}
In this section, we consider the problem of optimizing the deployment density
of a cellular network, given the transmit power of the base
stations, and aiming at energy efficiency optimization. By
assuming that the base stations are distributed according to
a Poisson point process, an accurate and realistic analytical
model for optimizing the energy efficiency has recently been
proposed in \cite{TWC_Paper2018}, and the optimal deployment density of
the cellular base stations has been formulated in a tractable analytical form. Leveraging \cite{TWC_Paper2018}, large datasets can be generated with low computational effort, which provides system
designers with the optimal base stations density as a function
of the base stations transmit power. It is known, on the other
hand, that similar tractable analytical frameworks cannot be
easily obtained if the cellular base stations are distributed
according to non-Poisson spatial models, which makes the
energy efficiency optimization of such cellular network deployments difficult \cite{nonPPP_TWC }. In addition, the generation of
large datasets based on non-Poisson point processes is a time
and memory consuming task, which makes it difficult to
obtain large datasets containing the optimal data pairs (optimal
deployment density, transmit power) for training purposes.

In the depicted context, thus, the two-step approach introduced and described in Section \ref{Sec:OptData} is well motivated.
We consider an ANN whose objective is to provide the
system designers with the optimal deployment density of the
base stations that optimizes the energy efficiency for a given
transmit power of the base stations. Thus, the transmit power
of the base stations is the input of the ANN and the optimal
deployment density of the base stations is the output of the
ANN. It is assumed, in particular, that the base stations are
distributed according to a non-Poisson point process, whose
exact distribution is not known, and that just (some) empirical
samples for the locations of the base stations are available.

We aim at understanding whether by first performing an
initial training of the ANN based on a large Poisson-based
dataset, and then executing a second training based on a small
dataset of empirical (or synthetic from simulations) data, we
can obtain similar performance as using only a large training
set of real data. To answer this question, we use the following
approach:

1) We generate a large dataset by assuming that the base
stations are distributed according to a Poisson point process,
and compute the optimal base station density as a function of
the transmit power by using the analytical framework recently
proposed in \cite{TWC_Paper2018}. An ANN is first trained by using the obtained dataset.

2) We generate a small dataset by considering the actual
(non-Poisson) spatial distribution for the locations of the base
stations, and compute the optimal base station density as a
function of the transmit power through an exhaustive search.
The pre-trained ANN obtained during the previous phase is
then refined by employing this second training set. In our implementation, the number of layers and neurons is determined
during the first training phase, and is left unchanged during the
second training phase. During the latter phase, in particular,
only the weight and bias coefficients are updated.

In the considered case study, more specifically, the actual
locations of the base stations are assumed to follow a square
grid model. This choice is motivated by the fact that the
Poisson-based spatial model provides one with a worst-case
approximation of the grid-based spatial distribution of the base
stations \cite{nonPPP_TWC }. Therefore, we test the suitability of the proposed
approach in the worst-case setup for the initial training.

\begin{figure}[!t]
  \begin{center}
  \includegraphics[width=\columnwidth]{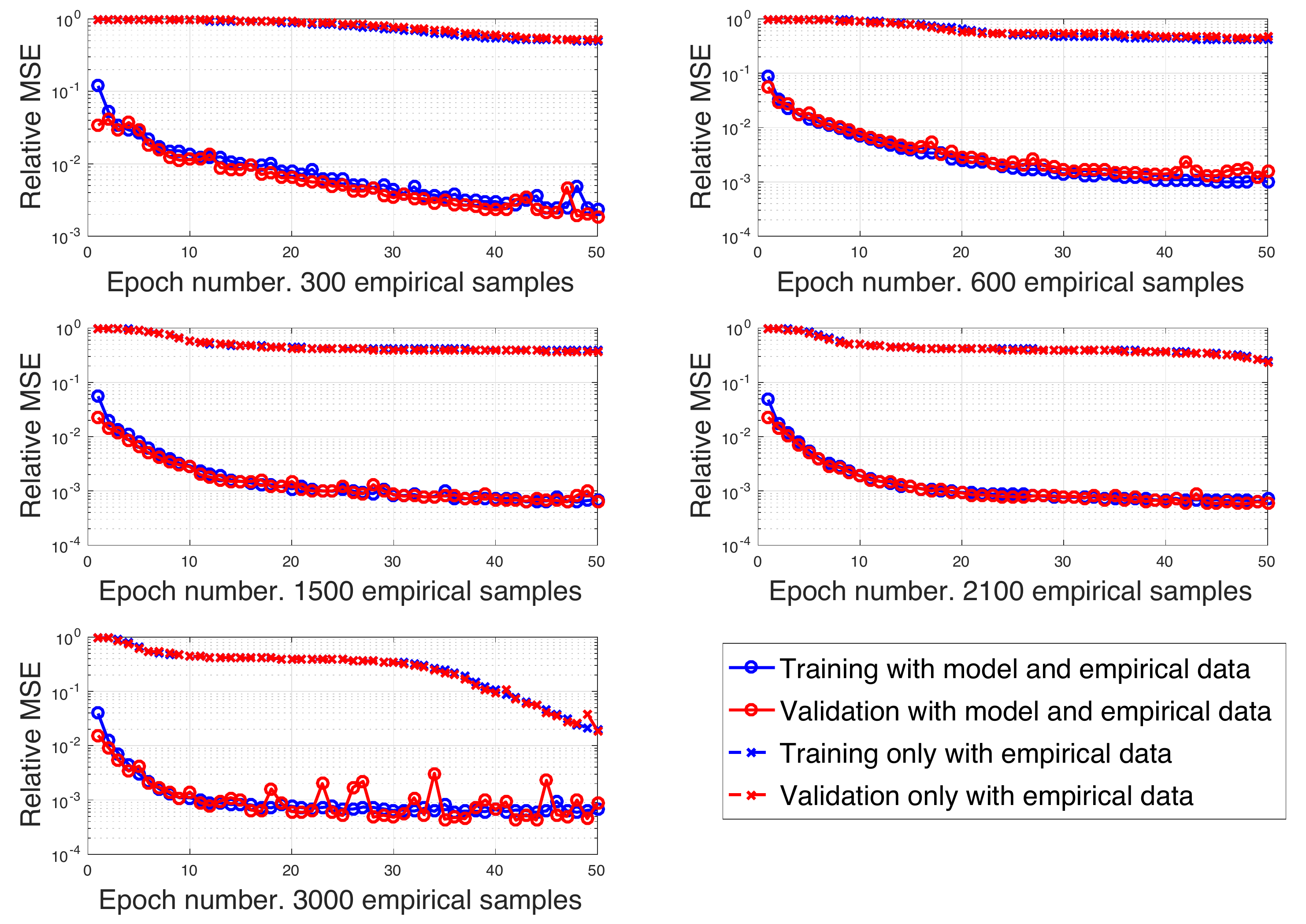}
  \caption{Learning and validation relative MSE vs. training epochs for x = 300, 600, 1500, 2100, and 3000 samples. For each case, the performance with and without PPP-based samples is reported. It is seen how the use of PPP-based data significantly improves the performance.}
  \label{Fig_nonPPP_Learning}
 \end{center}
\end{figure}

\begin{figure}[!t]
  \begin{center}
  \includegraphics[width=\columnwidth]{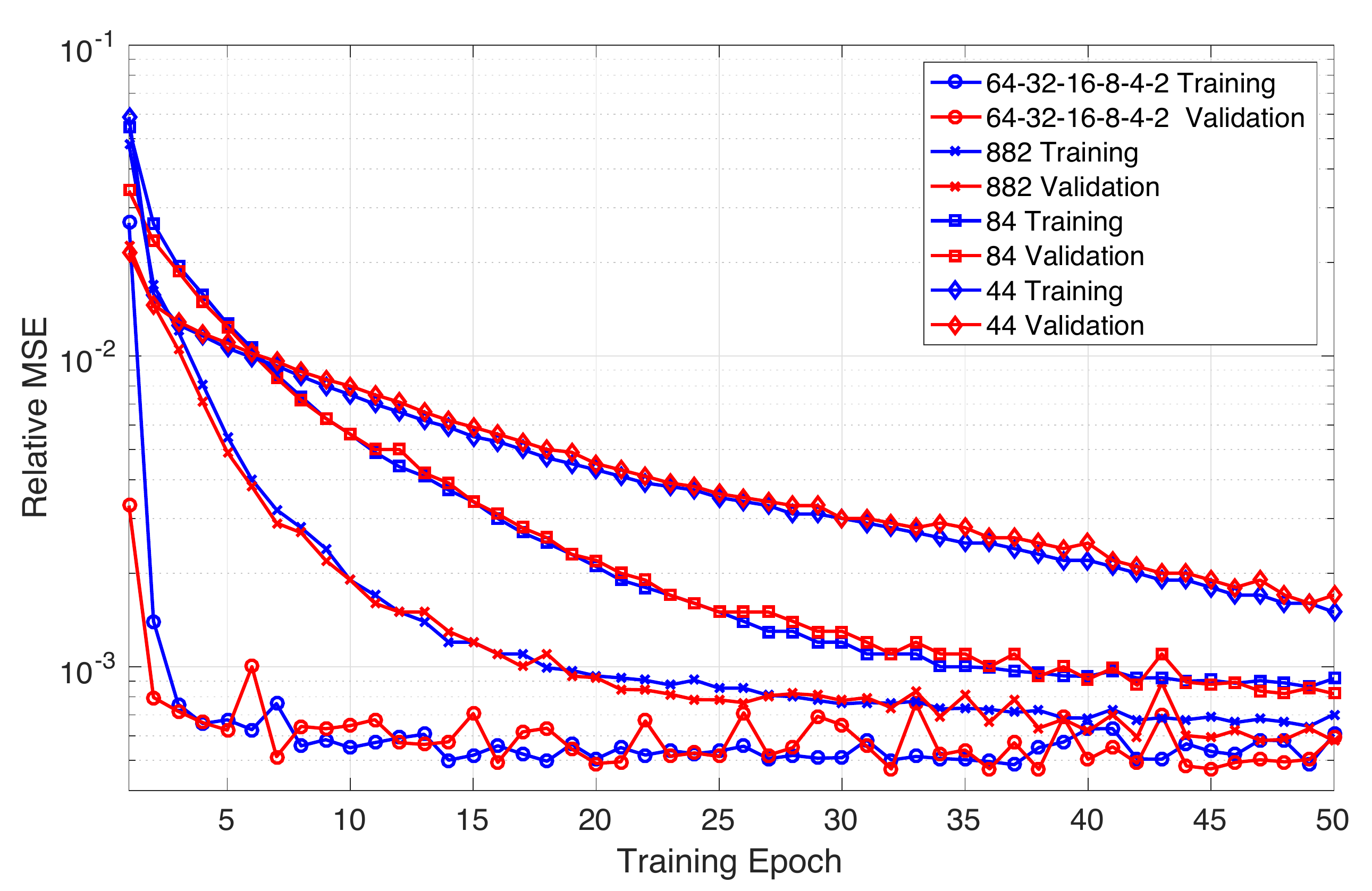}
  \caption{Learning and validation relative MSE vs. training epochs for different network architectures (case study x=2100 real samples).}
  \label{Fig_nonPPP_ANNarchitectures__1}
 \end{center}
\end{figure}

Figure \ref{Fig_nonPPP_Learning} shows the training and validation relative mean
square error (MSE) versus the number of training epochs for different amounts of empirical (real) data employed for training. A feed-forward ANN
architecture with fully-connected layers and ReLU activation
functions is considered \cite{Bengio2016}. Specifically, three hidden layers have been employed, with 8, 8, 2 neurons each and the following scenarios are analyzed:
1) the proposed approach, where model-based and empirical-based data samples are used, and 2) the baseline approach,
where only empirical-based data samples are used. As for the
first scenario, the size of the training set is always set equal
to 30,000 samples, out of which x samples follow the true
base station distribution (square grid model), and (30,000-x)
samples follow the Poisson distribution. The values x = 300,
600, 1500, 2100, and 3000 are reported. As for the second scenario, on the other hand, the size of the training set is set
equal to x empirical samples. The comparison between the two
case studies is, therefore, fair in terms of number of empirical
data samples employed and shows that augmenting a small dataset of empirical data with a larger dataset of model-based data can provide significant performance gains.

The choice of the ANN architecture considered in Fig. \ref{Fig_nonPPP_Learning} has been made after analyzing the performance of several ANN architectures with different numbers of neurons and layers. The results of this analysis are reported in Figure \ref{Fig_nonPPP_ANNarchitectures__1}, for the case x=2100 samples. They provide evidence that the chosen 8-8-2 configuration leads to lower training and validation errors than other ANN architectures of similar size, while exhibiting a small gap with respect to a more complex ANN with six hidden layers made of 64, 32, 16, 8, 4, 2 neurons, respectively. 


\begin{figure}[!t]
  \begin{center}
  \includegraphics[width=\columnwidth]{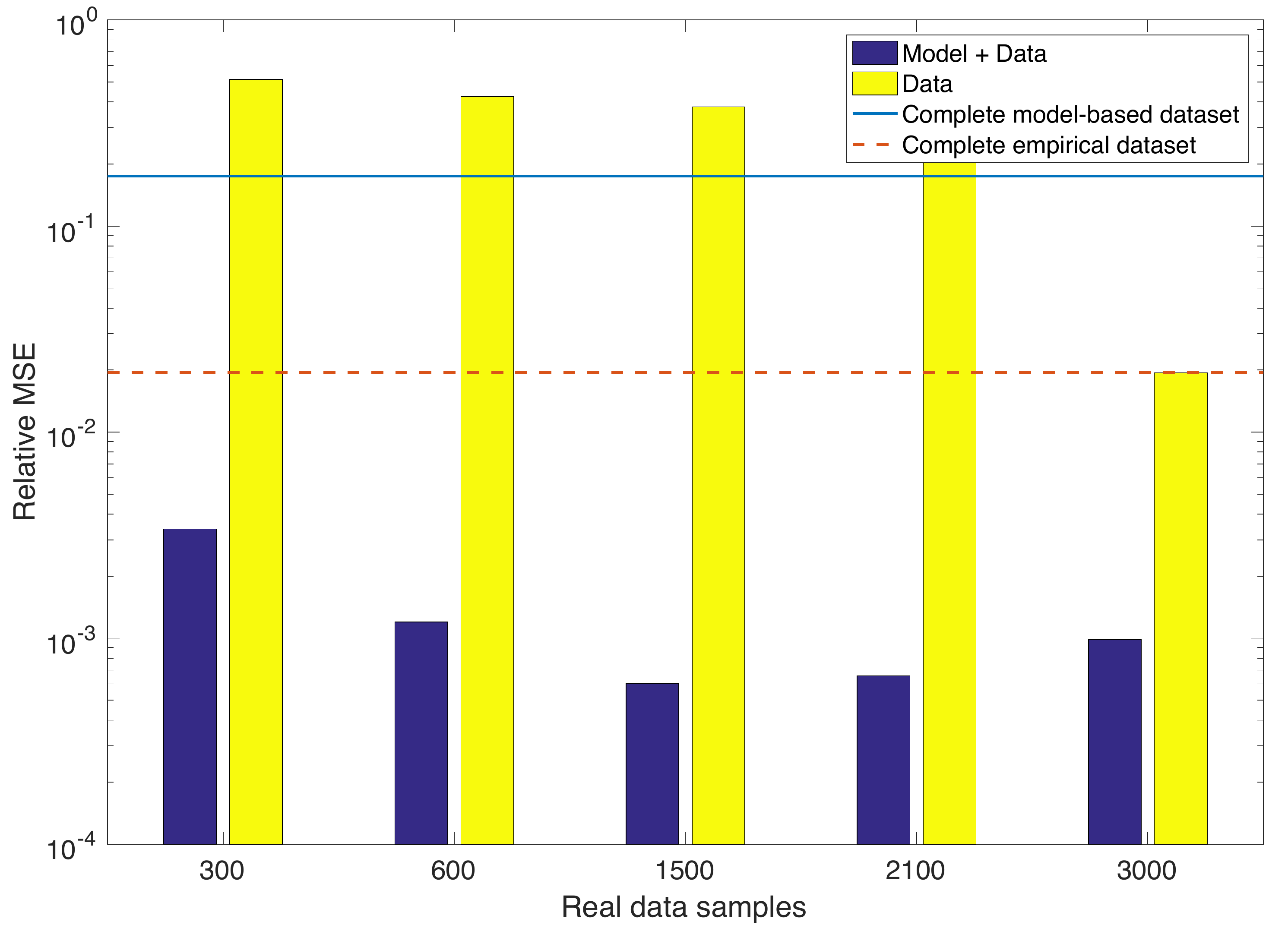}
  \caption{Test error vs. the number of real data samples used during training. For each case, the performance with and and without PPP-based samples is reported. The use of PPP-based data significantly improves the performance.}
  \label{Fig_nonPPP_TestError}
 \end{center}
\end{figure}

Finally, Fig. \ref{Fig_nonPPP_TestError} compares the test errors provided by the
proposed method and the pure data-driven approach. This figure
is obtained by considering the ANN network configuration
with three hidden layers having 8, 8, and 2 neurons respectively, and after 50 training epochs. The results indicate that our proposed approach performs better than pure data-driven methods that employ a small number of empirical data samples,
and outperforms even pure data-driven methods that use a
larger number of empirical data samples. The reason behind
this improved performance lies in the good initialization point
that model-based expert knowledge provides, i.e., a
more suitable (and closer-to-the optimal for a given number of
training epochs) initial configuration of the ANN. The figure
illustrates that using a pure model-based approach, on the
other hand, usually results in inaccurate system designs.

\subsection{Energy Efficiency Maximization with Unknown Power Consumption Models}
In this case study, we consider again the problem of optimizing the deployment density of a cellular network, given the transmit power of the base stations and by considering the energy efficiency as the key performance indicator of interest. In contrast to the previous case study, we assume that the cellular base stations are distributed according to a Poisson point process, which is considered to be accurate for the application of interest. We assume, however, that only a simplified statistical model for the static and idle hardware power consumptions of the cellular base stations  is available. Details about the definitions of static and idle power consumptions can be found in \cite{TWC_Paper2018}. Specifically, we consider that the static and idle hardware power consumptions are distributed according to two uniform random variables with some given mean and variance. Although tractable, this model is clearly a rough approximation of the actual hardware power consumption, which will in practice deviate from the considered model. Nevertheless, the numerical results that are shown in this section confirm that even such a simple model can provide us with enough prior information.

Based on the adopted uniform model, the optimal deployment density of the cellular base stations is computed, as a function of the transmit power and of the static and idle power consumptions, by employing the optimization framework proposed in \cite{TWC_Paper2018}, which allows us to easily produce large datasets for training ANNs. Subsequently, we generate a smaller dataset based on the actual realizations of the static and idle hardware power consumptions, which, for ease of reproducibility, are assumed to follow a Gaussian model with fixed mean and variance. We note that, as opposed to the case study in the previous section, the considered ANN has three inputs (the transmit power of the base stations, the static power consumption, the idle power consumption) and one output (the optimal deployment density of the base stations), and so it is more difficult to train as opposed to the one-input and one-output ANN considered in the previous case study.

With these two datasets available, we use a similar procedure as the one discussed in Section \ref{Sec:CaseII}:

1) First, the initial training of the ANN is performed based on the training set obtained from the approximate power consumption model (uniform distribution).

2) Then, the dataset obtained from the true values of the static and idle hardware power consumptions are used to refine the initial training.

\begin{figure}[!t]
  \begin{center}
  \includegraphics[width=\columnwidth]{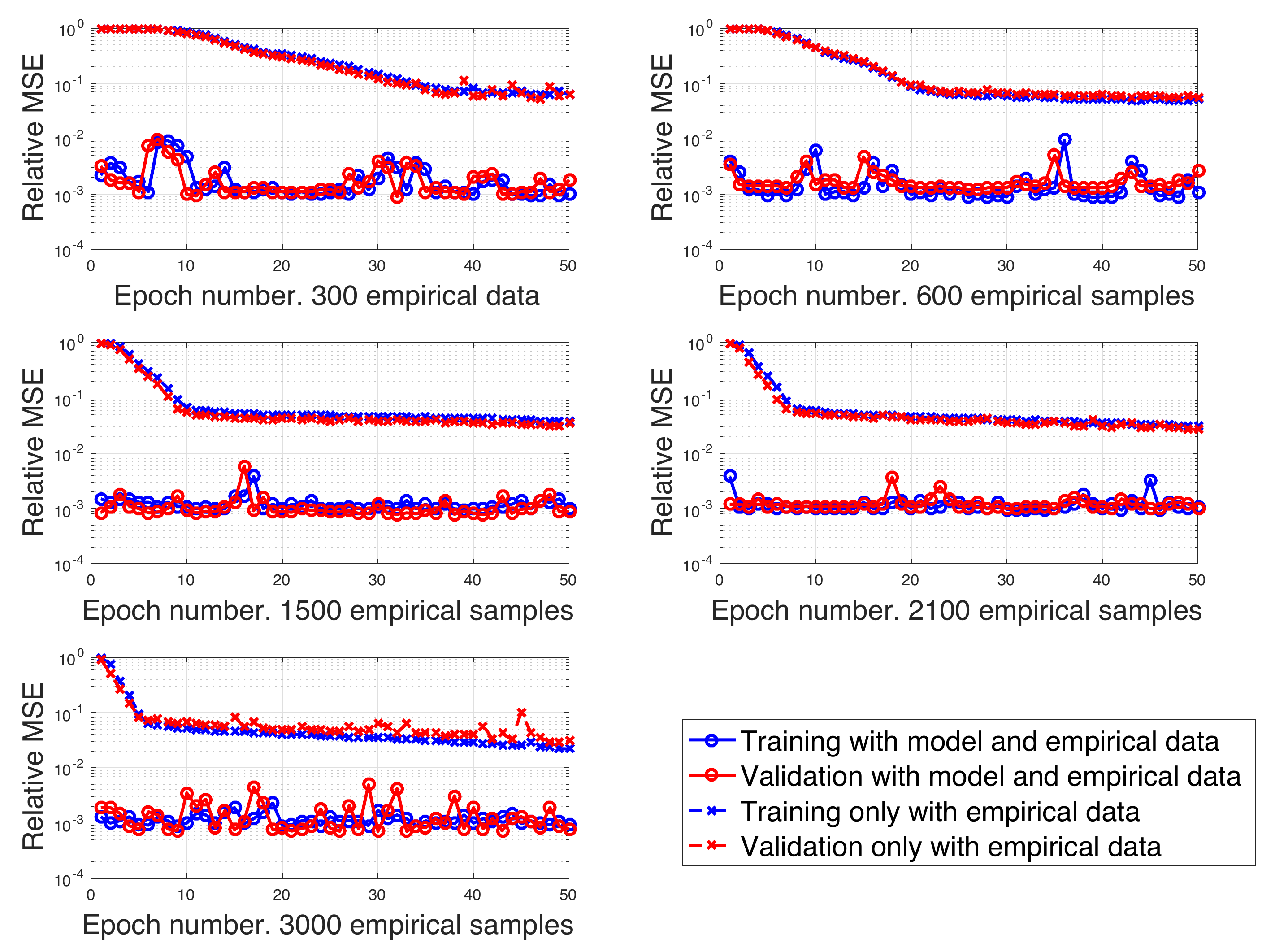}
  \caption{Learning and validation relative MSE vs. the training epochs for x=300, 600, 1500, 2100, and 3000. For each case, the performance with and and without PPP-based samples is reported. The use of PPP-based data significantly improves the performance.}
  \label{Fig_CircuitsPower_Learning}
 \end{center}
\end{figure}

\begin{figure}[!t]
  \begin{center}
  \includegraphics[width=\columnwidth]{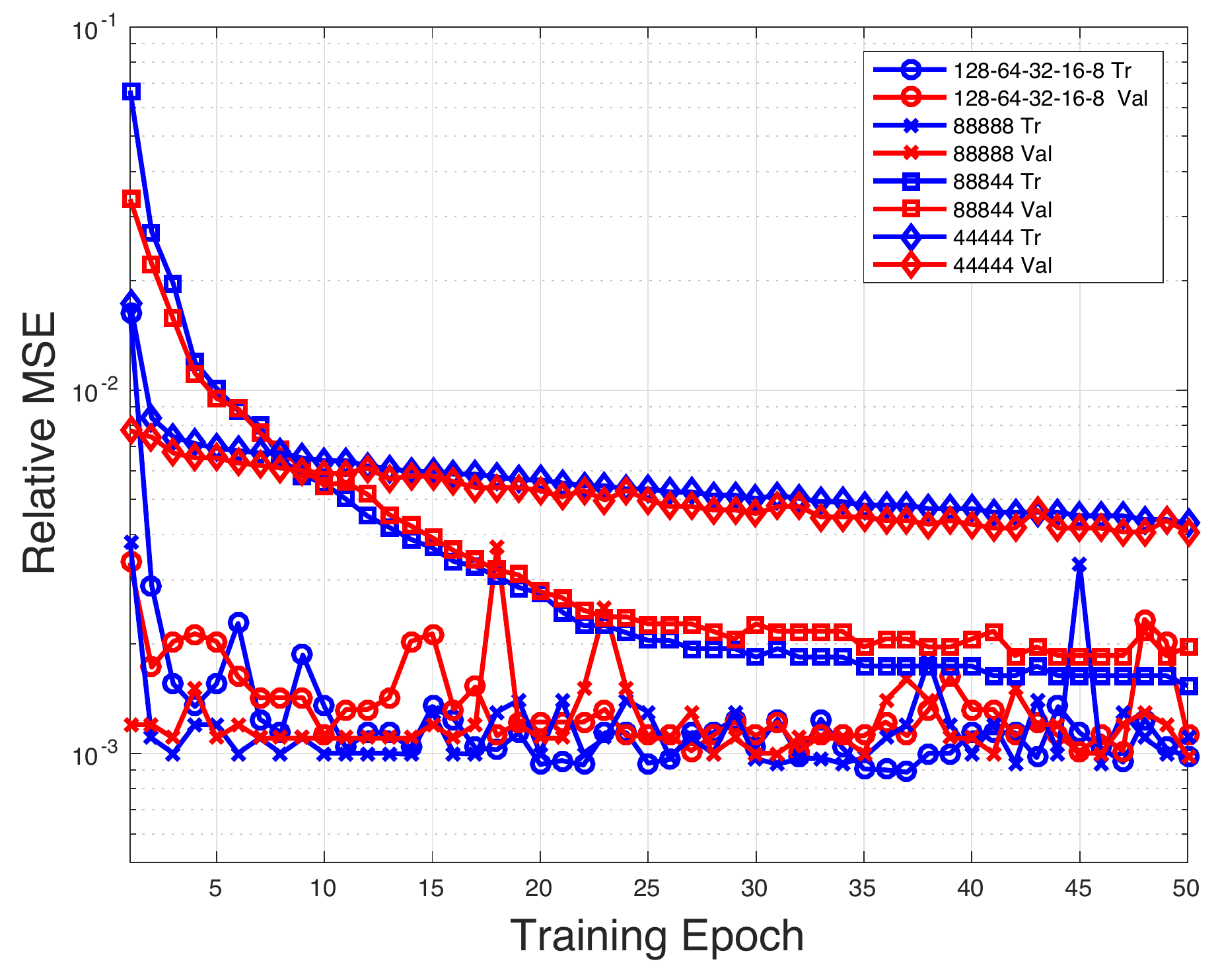}
  \caption{Learning and validation relative MSE vs. training epochs for different network architectures (case study x=2100).}
  \label{Fig_nonPPP_ANNarchitectures__2}
 \end{center}
\end{figure}

\begin{figure}[!t]
  \begin{center}
  \includegraphics[width=\columnwidth]{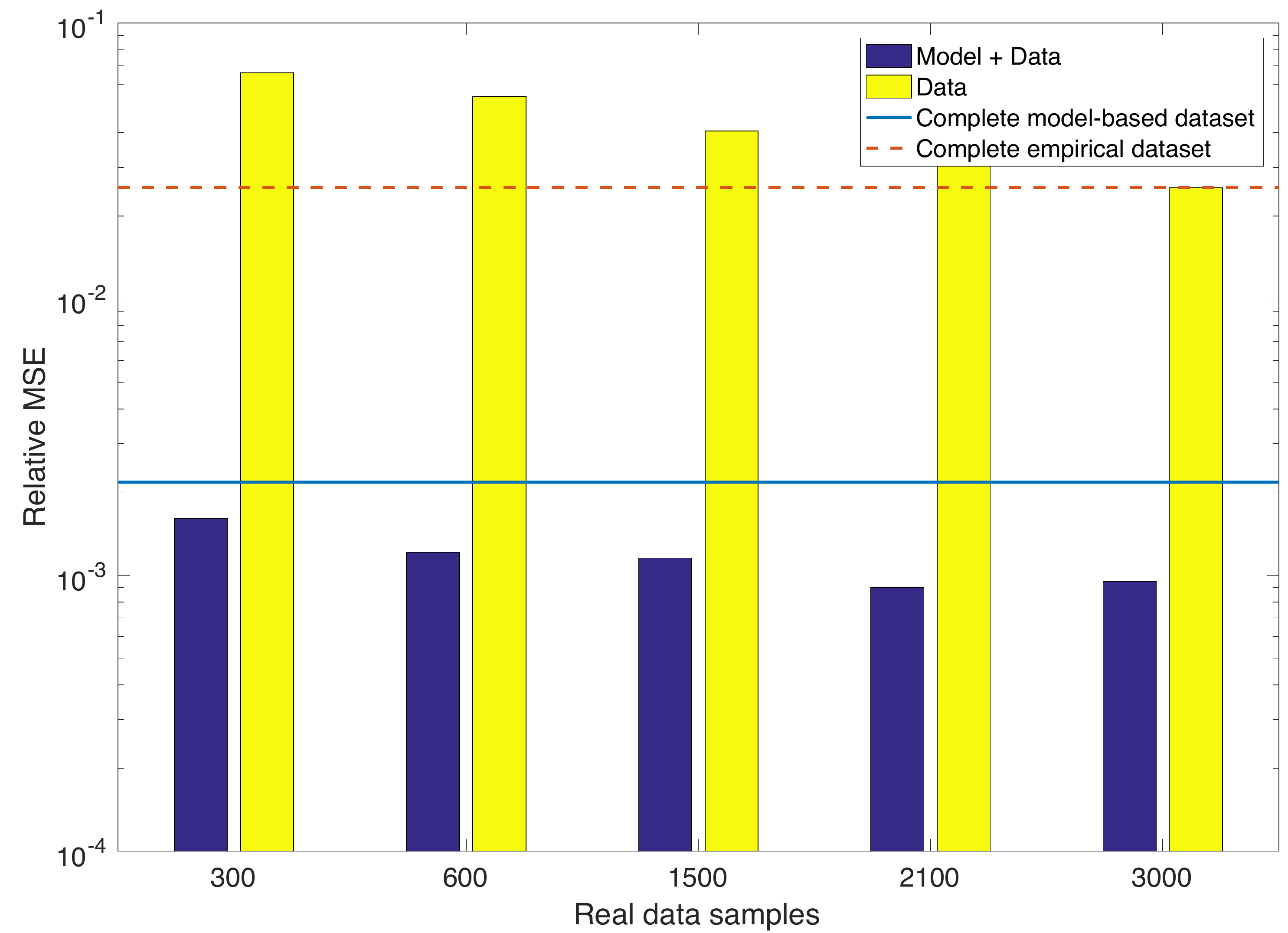}
  \caption{Test error vs. the number of real data samples used during training. For each case, the performance with and and without PPP-based samples is reported. The use of PPP-based data significantly improves the performance.}
  \label{Fig_CircuitsPower_TestError}
 \end{center}
\end{figure}

The results are illustrated in Figs. \ref{Fig_CircuitsPower_Learning}, \ref{Fig_nonPPP_ANNarchitectures__2}, and \ref{Fig_CircuitsPower_TestError} which are the counterparts of Figs. \ref{Fig_nonPPP_Learning}, \ref{Fig_nonPPP_ANNarchitectures__1} and \ref{Fig_nonPPP_TestError}, respectively. Specifically, Fig. \ref{Fig_nonPPP_ANNarchitectures__2} shows that the ANN configuration with the best complexity-performance trade-off is one with five hidden layers with 8, 8, 8, 8, and 8 neurons each, and ReLU activation functions. This ANN, remarkably, exhibits just slightly worse performance than an ANN with 128-64-32-16-8 neurons in the five hidden layers. For completeness, similar for the previous case study, the training, validation, and test error curves are reported in Figs. \ref{Fig_CircuitsPower_Learning} and \ref{Fig_CircuitsPower_TestError}, respectively. 


\section{Conclusion and Open Research Issues}
This work shows that mathematical models and optimization techniques provide unique insights to complement and improve ANN data-driven approaches. Unlike other application fields where deep learning may be employed, the solid theoretical understanding of wireless communication systems and networks that communication theorists have developed during the last decades provide us with unique opportunities to be exploited. Nevertheless, several important questions and open issues can be identified for future research. This includes the following.

1) How does the dimension of the training set scale up with the number of parameters to learn? It is anticipated that systems with a large number of parameters to be optimized will require more training data, but different scaling laws might be observed for different classes of optimization problems.

2) Is it possible to infer the resource allocation for a large system from the resource allocation of a smaller system? This would enable to scale up known ANN configurations to higher dimensions without  a new training phase.

3) When using a simplified model for pre-training, how is the accuracy of the simplified model related to the amount of training data to use in both phases? 

Finally, while the proposed approach represents a possible method to combine deep learning with mathematical modeling, it is to be mentioned that other approaches exist. More precisely, the proposed approach constitutes (and can be viewed as) an instance of (deep) \emph{transfer learning}, which studies how to transfer the knowledge that is used in a given context to execute a given task, into a different, but related context, to execute  another task. The approach described in Sections IV-B and IV-C is, more precisely, an instance of the so-called \emph{network-based transfer learning} method. In addition, the deep reinforcement learning framework \cite{Arulkumaran2017_DRL}, which is able to learn optimal action policies in a fully data-driven fashion from the direct interaction with the environment, is another promising approach to explore and to combine with transfer learning methods. 

\bibliographystyle{IEEEtran}
\bibliography{DeepLearning}

\end{document}